\documentclass[aps,prl,twocolumn]{revtex4}
\usepackage[pctex32]{graphics}

\begin{document}

\title{Implementation of controlled SWAP gates for quantum fingerprinting and
photonic quantum computation}
\author{B. Wang and L.-M. Duan}
\affiliation {FOCUS center and MCTP, Department of Physics,
University of Michigan, Ann Arbor, MI 48109}

\begin{abstract}
We propose a scheme to implement quantum controlled SWAP gates by
directing single-photon pulses to a two-sided cavity with a single
trapped atom. The resultant gates can be used to realize quantum
fingerprinting and universal photonic quantum computation. The
performance of the scheme is characterized under realistic
experimental noise with the requirements well within the reach of
the current technology.

\textbf{PACS numbers:} 03.67.-a, 42.50.Gy, 42.50.-p
\end{abstract}

\maketitle

Several quantum cryptographic schemes, such as quantum
fingerprinting \cite {1} and quantum digital signatures \cite{2},
require the controlled SWAP (CSWAP) gates as the critical element
for their realization. The qubits in these schemes are carried by
photon pulses as photons are the only viable choice for remote
quantum communication. The essential requirement
for implementation of these schemes is to measure the overlap of two $n$%
-qubit wave functions $\left| \Psi \right\rangle $ and $\left| \Psi ^{\prime
}\right\rangle $. A multi-qubit CSWAP gate provides the simplest method for
realization of this measurement \cite{1,2}. For a CSWAP gate, conditioned on
the state $\left| 1\right\rangle _{c}$ of the control qubit $c$, the two
sets of target qubits $A=\{1,2,\cdots ,n\}$ and $B=\{1^{\prime },2^{\prime
},\cdots ,n^{\prime }\}$ exchange their quantum states, so in general, we
have
\begin{eqnarray}
&&U_{\text{CSWAP}}\left( c_{0}\left| 0\right\rangle _{c}+c_{1}\left|
1\right\rangle _{c}\right) \otimes \left| \Psi \right\rangle _{A}\otimes
\left| \Psi ^{\prime }\right\rangle _{B}  \nonumber \\
&=&c_{0}\left| 0\right\rangle _{c}\otimes \left| \Psi \right\rangle
_{A}\otimes \left| \Psi ^{\prime }\right\rangle _{B}+c_{1}\left|
1\right\rangle _{c}\otimes \left| \Psi ^{\prime }\right\rangle _{A}\otimes
\left| \Psi \right\rangle _{B}.
\end{eqnarray}
If the control qubit is initially prepared in the state $\left( \left|
0\right\rangle _{c}+\left| 1\right\rangle _{c}\right) /\sqrt{2}$, and then
measured in the basis $\left\{ \left| \pm \right\rangle _{c}\equiv \left(
\left| 0\right\rangle _{c}\pm \left| 1\right\rangle _{c}\right) /\sqrt{2}%
\right\} $ after the CSWAP gate, the probability to get the outcome ``$-$''
is given by $p_{-}=1-\left| \langle \Psi |\Psi ^{\prime }\rangle \right|
^{2} $, which shows exactly the information about the state overlap $\left|
\left\langle \Psi |\Psi ^{\prime }\right\rangle \right| $. So the central
problem for implementation of these cryptographic schemes becomes how to
realize a multi-qubit CSWAP gate.

In this paper, we propose a scheme to realize a multi-qubit CSWAP
gate on two sequences of photon pulses $1,2,\cdots ,n$ and
$1^{\prime },2^{\prime },\cdots ,n^{\prime }$ by simply
transferring them through a single atom cavity. The cavity is
two-sided, and the pulse sequences are incident from the different
side mirrors. The atom inside the cavity plays the role of the
control qubit. Under an appropriate atomic level configuration,
this setup realizes exactly the CSWAP gate on the two pulse
sequences. Beyond its applications for implementation of the above
quantum cryptographic protocols, we also show that a simple
version of this CSWAP gate, together with single-bit polarization
rotations, realize universal quantum computation with photon
pulses as qubits. Compared with the recent proposal of the
photonic computation scheme based on the single-sided cavities
\cite{3}, this scheme has the advantage that it is directly built
on the state-of-the-art two-sided cavities \cite{4}. The same
scheme also applies to other experimental setups with optical
resonators, such as a quantum dot inside a solid state cavity
\cite{5,6,7}. To characterize performance of the CSWAP gate under
influence of realistic experimental noise, we provide detailed
theoretical modeling, and the calculation shows the practicality
of the gate under typical configurations of either the atomic or
the solid-state cavities. In particular, the scheme requires
neither the good cavity limit nor the Lamb-Dicke condition for the
trapped atom, which significantly simplifies its experimental
realization.

First, we explain the basic idea of our scheme for implementation of the
multi-qubit CSWAP gate. Consider an atom trapped in a two-sided optical
cavity with the relevant atomic levels shown in Fig.1. The cavity supports
two eigenmodes $a_{h}$ and $a_{v}$ with different polarizations (horizontal
``$h$'' and vertical ``$v$'', respectively). These two modes are resonantly
coupled to the corresponding atomic transitions $|0\rangle \leftrightarrow
|e_{h}\rangle $ and $|0\rangle \leftrightarrow |e_{v}\rangle $ ($%
|e_{h}\rangle $ and $|e_{v}\rangle $ could be superpositions of the Zeeman
states on the same excited hyperfine manifold). The state $|1\rangle $ is on
a different hyperfine level in the ground-state manifold, and is decoupled
from the cavity modes due to the large hyperfine splitting. The two cavity
modes $a_{h}$ and $a_{v}$ are resonantly driven by the ``$h$'' and ``$v$''
polarization components of the single-photon pulses incident on the cavity
mirrors, respectively. Each single-photon pulse represents an optical qubit,
with its qubit basis-state carried by the polarization ``$h$'' or ``$v$''.

If the atom is prepared in the state $|1\rangle $, the input pulses
basically see an empty cavity as the atom is decoupled from the cavity
modes. With such a resonant cavity, the input pulses from both sides will
directly go through if their bandwidth is significantly smaller than the
cavity decay rate $\kappa $ (assuming that both mirrors of the cavity give
rise to the same decay rate). The states of the two pulse sequences from
different sides are exchanged. (See Fig.1 for convention of the notation. We
assume that the pulses from different sides have the same pulse shape.)
However, if the atom is prepared in the state $|0\rangle $, due to the
strong atom-cavity coupling, the transmission spectrum of the dressed cavity
is significantly modified, and the pulses from both sides will be reflected
by the cavity mirrors if their bandwidth is significantly smaller than $%
g^{2}/\kappa $, where $g$ is the atom-cavity coupling rate (see the
following theoretical modeling). The states of the pulses remain unchanged
\cite{note1}. From consideration of the above two cases, we see that if the
atom is prepared in a superposition of the states $|0\rangle $ and $%
|1\rangle $, this cavity setup performs exactly the CSWAP gate described by
Eq. (1), with the atom as the control qubit and the two $n$-qubit
pulse-sequences on different side mirrors as the target qubits.
\begin{figure}[tbp]
\includegraphics{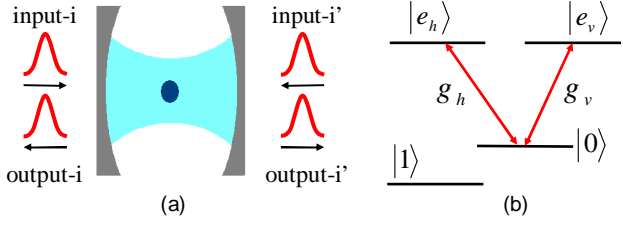}
\caption[Fig.1]{(a) Schematic setup for implementation of the
multi-qubit CSWAP gate. Two pulse sequences $i, (i=1,2\cdots n)$
and $i^{\prime }, (i^{\prime }=1^{\prime },2^{\prime }\cdots
n^{\prime })$ are incident on and then reflected from (or
transmitted through) the single-atom cavity. (b) Configuration of
the relevant atomic levels.}
\end{figure}

As mentioned before, the multi-qubit CSWAP gate implemented by this cavity
setup is ideal for the measurement of overlap of two $n$-qubit wave
functions carried by the photon pulses. Therefore, the scheme is critical
for realization of a number of quantum cryptographic protocols, including
quantum fingerprinting and quantum digital signature \cite{1,2}. Beyond this
important application, here we also want to show that the simplest version
of this gate, the CSWAP on two optical qubits (denoted as CSWAP$_{2}$), also
provides a critical gate, which, together with simple single-qubit
rotations, realize universal quantum computation. In this computational
scheme, the qubits are represented by the single-photon pulses, which have
the advantages of being relatively easy to scale to many qubits and to
integrate into quantum networks. The atom (or the quantum dot in the
solid-state cavity) only acts as an ancilla qubit which mediates strong
interaction between the photons during the gate operation.

To see the universality of the CSWAP$_{2}$ gate, it is enough to show that,
together with single-bit rotations, it leads to the standard controlled
phase flip (CPF) gate on two arbitrary photonic qubits. In Fig. 2(a), from
the CSWAP$_{2}$ gates we give one construction of the CPF\ gate $%
U_{CPF}=e^{i\pi |hv\rangle _{12}\left\langle hv\right| }$, which flips the
phase of the photons 1 and 2 if and only if they are in the state component $%
|hv\rangle $. The atomic qubit is initially prepared in the state $|\varphi
\rangle _{a}=(|0\rangle +|1\rangle )/\sqrt{2}$, which is recovered after the
whole operation. In this construction, we use four CSWAP$_{2}$ gates,
together with a few single-bit Hadamard gates H and $i$-phase gates (the
latter adds a phase $i$ to the state component $|1\rangle $). This
construction can be further simplified if we use feed-forward from a
measurement on the atomic qubit. In Fig. 2b, we give a simplified circuit
which uses only two CSWAP$_{2}$ gates. After the operation represented by
the left side of this circuit, we perform a measurement on the atomic qubit
in the basis $\left\{ |0\rangle ,|1\rangle \right\} $, and upon the outcome
``$1$'', we add a $\sigma _{z}=|h\rangle \langle h|-|v\rangle \langle v|$ ($Z
$-gate) to each of the photonic qubits. One can verify that the whole
operation also performs the gate $U_{CPF}$ on the two photonic qubits. These
constructions prove that the CSWAP$_{2}$ gates, combined with single-bit
rotations, indeed realize universal quantum computation on photon pulses.

\begin{figure}[tbp]
\includegraphics {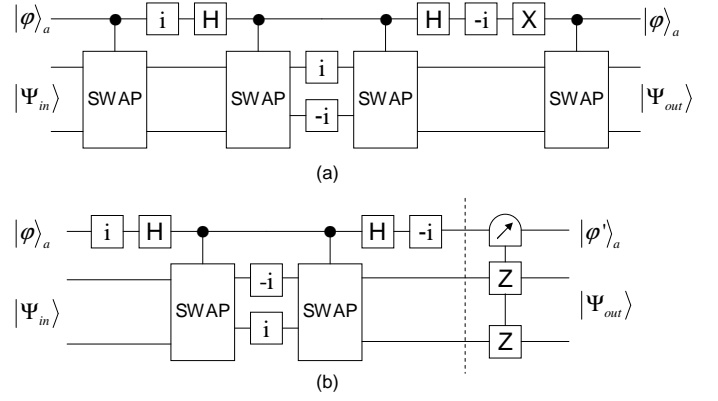}
\caption[Fig.2]{(a)A circuit to construct a controlled phase flip
gate from the CSWAP gates. (b)An alternative circuit which uses
only two CSWAP gates, but with feed-forward of a measurement on
the atomic qubit (the right side of the dashed line, see the text
for explanation).}
\end{figure}

Now we proceed to present the detailed theoretical modeling of interaction
between photonic pulses and an atom inside a two-sided cavity. This
calculation serves for two purposes; first, we need to prove the statements
made before for the principle of the CSWAP gate. In particular, we will show
that the pulse will have reflection or transmission conditional on the
atomic state if its bandwidth is much smaller than the coupling rates $%
\kappa $ and $g^{2}/\kappa $. Second, we also want to characterize the
influence of noise on this gate scheme. The most important noise here is the
intrinsic atomic spontaneous emission, which causes loss of photons to
uncontrolled directions. There could be other source of noise, such as the
light absorption/scattering at the cavity mirrors. The effect of the latter,
however, is very similar to the atomic spontaneous emission and thus can be
similarly modeled.

The interaction between the atom (or, in general, the dipole) and the cavity
modes is described by the Hamiltonian in the rotating frame (see Fig. 1 for
the notations)
\begin{equation}
H=\sum_{\mu =h,v}g_{\mu }\left( \sigma _{\mu }^{+}a_{\mu }+\sigma _{\mu
}^{-}a_{\mu }^{\dagger }\right) ,
\end{equation}
where $\sigma _{\mu }^{+}\equiv |e_{\mu }\rangle \langle 0|$ is the atomic
raising operator, $\mu =h,v$ denote the polarization modes, and $g_{_{\mu }}$
are the corresponding atom-cavity coupling rates. The cavity modes $a_{\mu }$
are driven by the corresponding input fields $a_{\mu ,l}^{in}$ and $a_{\mu
,r}^{in}$ from both the left and the right sides of the cavity. The
Heisenberg-Langevin equations for $a_{\mu }$ have the form \cite{8}
\begin{eqnarray}
\dot{a}_{\mu } &=&-ig_{\mu }\sigma _{\mu }^{-}-\left( \kappa _{\mu
,l}+\kappa _{\mu ,r}\right) a_{\mu }/2  \nonumber \\
&+&\sqrt{\kappa _{\mu ,l}}a_{\mu ,l}^{in}+\sqrt{\kappa _{\mu ,r}}a_{\mu
,r}^{in},
\end{eqnarray}
where $\kappa _{\mu ,l}$ and $\kappa _{\mu ,r}$ are the corresponding cavity
decay rates. The output fields $a_{\mu ,j}^{out}$ ($\mu =h,v$ and $j=l,r$)
are connected with the input fields through the cavity input-output relation
\cite{8}
\begin{equation}
a_{\mu ,j}^{out}=-a_{\mu ,j}^{in}+\sqrt{\kappa _{\mu ,j}}a_{\mu }.
\end{equation}
Both the input and the output fields satisfy the standard commutation
relations $\left[ a_{\mu ,j}^{in}(t),a_{\mu ^{\prime },j^{\prime
}}^{in\dagger }(t^{\prime })\right] =\left[ a_{\mu ,j}^{out}(t),a_{\mu
^{\prime },j^{\prime }}^{out\dagger }(t^{\prime })\right] =\delta _{\mu \mu
^{\prime }}\delta _{jj^{\prime }}\delta (t-t^{\prime })$. To complete the
set of equations, we also need the Heisenberg-Langevin equations for the
atomic operators, which have the form
\begin{equation}
\dot{\sigma _{\mu }^{-}}=-i\left[ \sigma _{\mu }^{-},H\right] -\gamma _{\mu
}\sigma _{\mu }^{-}/2+\sqrt{\gamma _{\mu }}\sigma _{\mu }^{z}\hat{N}_{\mu },
\end{equation}
where $\gamma _{\mu }$ denotes the spontaneous emission rate of the atomic
level $|e_{\mu }\rangle $, $\sigma _{\mu }^{z}\equiv |e_{\mu }\rangle
\langle e_{\mu }|-|0\rangle \langle 0|$, and $\hat{N}_{\mu }$ is the
corresponding vacuum noise operator which helps to preserve the desired
commutation relations for the atomic operators.

To characterize the CSWAP gate operation, we need to know the cavity output
fields given the inputs. Equations (3)-(5) completely determine the dynamics
of the system, but in general it is hard to solve this set of nonlinear
operator equations. However, we note in this scheme the atom has a rare
opportunity to stay in the excited states $|e_{\mu }\rangle $, so the matrix
elements for the components $|e_{\mu }\rangle \langle e_{\mu ^{\prime }}|$
should be negligible. We have done some exact numerical simulation with the
method specified in Refs. \cite{3,9}, which also confirms this
approximation. Under this approximation, $-\sigma _{\mu }^{z}$ is replaced
by the state projector $P_{0}=|0\rangle \langle 0|$, and the set of
equations (3)-(5) become linearized \cite{6,10}. The linearized equations
can be easily solved analytically by taking the Fourier transforms, and the
output fields are specified by the solution
\begin{equation}
a_{\mu ,j}^{out}(\omega )=R_{\mu }(\omega )a_{\mu ,j}^{in}(\omega )+T_{\mu
}(\omega )a_{\mu ,\bar{j}}^{in}(\omega )+m_{\mu }(\omega )\hat{N}_{\mu
}(\omega )\text{,}
\end{equation}
where $\{j,\bar{j}\}\equiv \{l,r\}$ or $\{r,l\}$, and for simplicity we have
taken $\kappa _{\mu ,r}=\kappa _{\mu ,l}\equiv \kappa _{\mu }$. The
operators $a_{\mu ,j}^{in}(\omega )$ and $a_{\mu ,j}^{out}(\omega )$ denote
the Fourier transforms of the input and the output field operators $a_{\mu
,j}^{in}(t),a_{\mu ,j}^{out}(t)$ with respect to time $t$. The reflection,
the transmission, and the noise coefficients $R_{\mu }(\omega ),T_{\mu
}(\omega )$, and $m_{\mu }(\omega )$ are given respectively by

\begin{eqnarray}
R_{\mu }(\omega ) &=&\frac{i\omega +g_{\mu }^{2}P_{0}/\left( i\omega -\gamma
_{\mu }/2\right) }{\kappa _{\mu }-i\omega -g_{\mu }^{2}P_{0}/\left( i\omega
-\gamma _{\mu }/2\right) }, \\
T_{\mu }(\omega ) &=&\frac{\kappa _{\mu }}{\kappa _{\mu }-i\omega -g_{\mu
}^{2}P_{0}/\left( i\omega -\gamma _{\mu }/2\right) }, \\
m_{\mu }(\omega ) &=&\frac{i\sqrt{\kappa _{\mu }\gamma _{\mu }}g_{\mu
}P_{0}/\left( i\omega -\gamma _{\mu }/2\right) }{\kappa _{\mu }-i\omega
-g_{\mu }^{2}P_{0}/\left( i\omega -\gamma _{\mu }/2\right) }.
\end{eqnarray}
From these expressions, we see that if the pulse bandwidth $\delta
\omega $ (the range of $\omega $) is much smaller than the rates
$\kappa _{\mu }$ and $g_{\mu }^{2}/\kappa _{\mu }$, and the noise
satisfies the condition $\gamma
_{\mu }\ll g_{\mu }^{2}/\kappa _{\mu }$, we have the reflection coefficient $%
R_{\mu }(\omega )\simeq -1$ for the atom in the state $|0\rangle $ and the
transmission coefficient $T_{\mu }(\omega )\simeq 1$ for the atom in the
state $|1\rangle $. This exactly confirms the statements that we used for
establishment of the CSWAP gate.

To quantitatively characterize the performance of the gate, we
need to specify the evolution from the input state to the output
state. The atom is assumed to be initially in the state $|\varphi
\rangle _{a}=(|0\rangle +|1\rangle )/\sqrt{2}$. The input state of
the two single-photon pulses from two sides of the cavity can be
expressed as $|\Psi \rangle _{p}=\sum_{\mu ,\mu ^{\prime }}C_{\mu
\mu ^{\prime }}|\mu \rangle _{l}|\mu ^{\prime }\rangle _{r}$,
where $\mu ,\mu ^{\prime }=h,v$. The qubit basis state $|\mu
\rangle
_{j}$ for a single-photon pulse is connected with the input field operator $%
a_{\mu ,j}^{in}(\omega )$ through $|\mu \rangle _{j}=\int f(\omega )a_{\mu
,j}^{in\dagger }(\omega )d\omega |vac\rangle $, where $|vac\rangle $ denotes
the vacuum state, and $f(\omega )$ is the normalized pulse shape function in
the frequency domain (which has been assumed to be the same for all the
input pulses). The state of the output pulses has a similar form but with $%
a_{\mu ,j}^{out\dagger }(\omega )$ replacing $a_{\mu ,j}^{in\dagger }(\omega
)$. As the expression of the output field operator $a_{\mu ,j}^{out}(\omega )
$ in Eq. (6) depends on the atomic projector $P_{0}$, the output state of
the photons gets entangled with the atomic state, as one expects for the
CSWAP gate.

We can use two quantities to characterize the gate performance: first, due
to the atomic spontaneous emission, we could lose a photon during the gate
operation with one of the output modes going to the vacuum state. So we use
the loss probability $p$ to characterize the inefficiency of the gate
operation. Second, even if both photons show up in the output, their pulse
shapes will be slightly distorted due to the frequency-dependent reflection
and transmission of the finite bandwidth input pulses. We can use the
fidelity to characterize the effect of this pulse shape distortion. To be
more specific, we consider a typical initial state $|\Psi _{in}\rangle
=|\varphi \rangle _{a}\otimes |hv\rangle _{l,r}$ for the atom and the
photons. In the ideal case, we should get the entangled output state $|\Psi
_{out}\rangle =\left( |0\rangle |hv\rangle _{l,r}+|1\rangle |vh\rangle
_{l,r}\right) /\sqrt{2}$, but in real case we in general get a density
matrix $\rho _{out}$ after tracing over the noise operator. The overlap $%
F\equiv $ $\left\langle \Psi _{out}\right| \rho _{out}|\Psi _{out}\rangle $
defines the fidelity, and we use it to characterize the gate performance
\cite{note2}. In the above characterization, we distinguish the inefficiency
and the infidelity errors for the gate, as the dominant error in this scheme
is the inefficiency error which allows for efficient quantum error
correction \cite{11}.

From the solution of $a_{\mu ,j}^{out}(\omega )$ in Eq.(6) and its
connection with the output state, we can calculate the loss
probability $p$\
and the fidelity $F$ as defined above. Their expressions are given by $%
p=1-\left( t_{h}^{(1)}t_{v}^{(1)}+r_{h}^{(0)}r_{v}^{(0)}\right) /2$ and $%
F=\left| \xi _{h}^{(0)}\xi _{v}^{(0)}+\xi _{h}^{(1)}\xi
_{v}^{(1)}\right| ^{2}/4$, where the superscripts $(0)$ and $(1)$
denotes the corresponding atomic state, $t_{\mu }^{(1)}=\int
d\omega \left| f\left( \omega \right) T_{\mu }^{(1)}(\omega
)\right| ^{2}$, $r_{\mu }^{(0)}=\int d\omega \left| f\left( \omega
\right) R_{\mu }^{(0)}(\omega )\right| ^{2}$, $\xi _{\mu
}^{(1)}=\int d\omega \left| f(\omega )\right| ^{2}T_{\mu }^{(1)}(\omega )/%
\sqrt{t_{\mu }^{(1)}}$, $\xi _{\mu }^{(0)}=\int d\omega \left|
f(\omega )\right| ^{2}R_{\mu }^{(0)}(\omega )/\sqrt{r_{\mu
}^{(0)}}$. We take the pulse shape $f(\omega )$ to be a Gaussian
function in the form $f(\omega )=\exp (-\omega ^{2}/\delta \omega
^{2})/\left( \sqrt{\pi }\delta \omega \right) $ with a bandwidth
$\delta \omega $. From these expressions, we calculate the loss
probability $p$ and the fidelity $F$ as functions of the scaled
pulse bandwidth and the atom-cavity coupling rate. The results are
shown in Fig. 3.

\begin{figure}[tbp]
\includegraphics {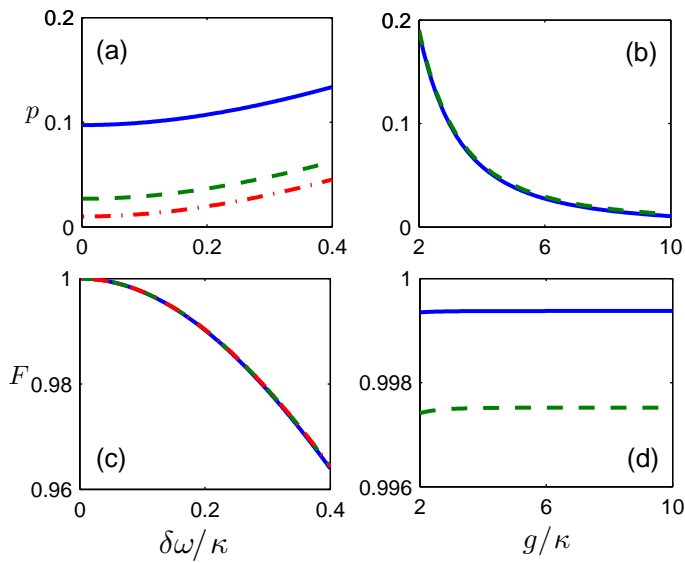}
\caption[Fig.3]{Left column: The loss probability $p$ and the
fidelity $F$ as a function of the scaled bandwidth
$\delta\omega/\kappa$, with $g=3\protect\kappa $ for the solid
line, $g=6\protect\kappa $ for the dashed line, and
$g=10\protect\kappa $ for the dash-dotted line. Right column: The
loss $p$ and the fidelity $F$ as a function of the scaled
atom-cavity coupling rate $g/\kappa$, with $\protect\delta
\protect\omega =0.05\protect\kappa $ for the solid line and
$\protect\delta \protect\omega =0.1\protect\kappa $ for the dashed
line. In the whole figure, we take $\kappa _{h}=\kappa _{v}=\kappa
$, $\gamma _{h}=\gamma _{v}=\gamma $, and $\gamma =\kappa $ for
simplicity.}
\end{figure}

Several remarks are in order from this calculation. First, the
fidelity $F$ is basically independent of the coupling rate $g$.
The loss probability $p$ depends on $g$, but $p$ remains small as
long as the variation in $g$ does not reduce $g$ close to zero.
This shows that the scheme here allows random variation of the
coupling rate $g$ in a significant range. That is a valuable
feature for the atomic cavity, as the thermal motion of the atom
typically brings it outside of the Lamb-Dicke limit which induces
significant random variation of the coupling rate $g$ in current
experiments
\cite{4}. Second, the scheme here does not require the good cavity limit $%
g>\kappa $. Independent of the ratio $g/\kappa $, the loss remains
small as long as we have the strong coupling condition
$g^{2}/\kappa \gamma \gg 1$ (or called the Purcell condition).
This is a valuable feature for the solid state cavity as it is
typically hard to get $g>\kappa $ for this setup although the
Purcell condition $g^{2}/\kappa \gamma \gg 1$ can be satisfied
\cite{6,7}. Third, we note that in this scheme the gate infidelity
is basically set by the finite pulse bandwidth $\delta \omega
/\kappa $, which in principle can be arbitrarily reduced. The
intrinsic noise, such as the atomic spontaneous emission (or other
kinds of photon loss) only leads to the gate inefficiency errors.
Finally, as some explicit parameter estimation, we have the
fidelity $F=99.75\%$ and the loss $p=1.3\%$ with the parameters
$(g,\kappa ,\gamma )/2\pi =(32,4.2,2.6)$ MHz and\ $\delta \omega
=0.1\kappa $, as typical for the atomic cavity \cite{4}; and $F=99.76\%,$ $%
p=1.59\%$ with $(g,\kappa ,\gamma )/2\pi =(0.66,6,0.001)$ THz and
$\delta \omega =0.1g^{2}/\kappa $, as typical for a solid state
cavity \cite{6,7}.

In summary, we have proposed a scheme to realize multi-qubit
controlled SWAP gates, which have critical application for
implementation of both quantum cryptographic protocols and
photonic quantum computation. The scheme has a number of nice
features that make it robust to practical noise and realizable
with the current technology.

We thank Jeff Kimble for helpful discussion. This work was supported by the
ARDA under ARO contracts, the NSF award (0431476), and the A. P. Sloan
Fellowship.


\begin{thebibliography}{99}
\bibitem{note1}  There is actually a sign flip of the atomic state $\left|
0\right\rangle $ with reflection of each pulse, but this sign flip can be
easily compensated afterwards by a $\sigma _{z}$ operation on the atom.
\bibitem{note2}  The fidelity in general depends on the input state, but for
different inputs it is of the same order of magnitude.
\end{thebibliography}
\end{document}